\documentclass[showpacs]{revtex4}
\usepackage[colorlinks=true,linkcolor=blue,citecolor=blue]{hyperref}

\usepackage[dvips]{graphicx}
\usepackage{epsfig}
\usepackage{ulem}
\usepackage[T1]{fontenc}
\usepackage[latin9]{inputenc}
\usepackage{esint}
\usepackage[english]{babel}
\usepackage{ulem}
\usepackage{color}
\makeatletter

\begin{document}

\title{On localized vegetation patterns,  fairy circles and localized patches in arid landscapes}

\author{ D. Escaff$^{1}$, C. Fernandez-Oto$^{2}$, M.G. Clerc$^{3}$ and M. Tlidi$^{2}$}

\affiliation{$^{1}$Complex Systems Group, Facultad de Ingenier{\'i}a y Ciencias
Aplicadas, Universidad de los Andes, 
Monse{\~n}or Alvaro del Portillo  12.455, Las Condes, Santiago, Chile,}
\affiliation{$^{2}$Facult{\'e} des Sciences, Universit{\'e} Libre de Bruxelles (U.L.B.), C.P.
231, Campus Plaine, B-1050 Bruxelles, Belgium}
\affiliation{$^{3}$Departamento de F{\'i}sica, Universidad de Chile, Blanco
Encalada 2008, Santiago, Chile.}

\begin{abstract}
We investigate the formation of localized structures with a varying width in one and   
two-dimensional systems. The mechanism of stabilization is attributed to strong nonlocal coupling 
mediated by a Lorentzian type of Kernel. 
We show that, in addition to stable dips  found recently [see, e.g., C. Fernandez-Oto, M. G. Clerc, D. Escaff, and M. Tlidi, Phys. Rev. Lett. {\bf{110}}, 174101 (2013)], exist stable localized  peaks which appear as a result of strong nonlocal coupling, i.e. mediated by a coupling that decays 
with the distance slower than an exponential. 
We applied this mechanism to arid ecosystems by considering a prototype model of a Nagumo type.  
In one-dimension, we study the front that connects the stable uniformly vegetated  state with the bare one under the effect of strong nonlocal coupling. We show that strong nonlocal coupling stabilizes 
both---dip and peak---localized structures. We show analytically and numerically that  
the width of localized dip, which we interpret as fairy circle, increases strongly with the aridity parameter. 
This prediction is in agreement with filed observations. In addition, we predict that the width of localized patch
decreases with the degree of aridity.  
Numerical results are in close agreement with analytical predictions.
\end{abstract}
\maketitle
\section{Introduction}
Localized structures (LS's) in dissipative media have been observed in various field of 
nonlinear science such as fluid dynamics, optics, laser physics, chemistry, and plant ecology 
(see recent  overviews \cite{Leblond-Mihalache,Tlidi-PTRA}).  
Localized structures consist of isolated or randomly distributed spots surrounded by regions in the uniform  state. 
They may consist of dips embedded in the homogeneous background. 
They are often called spatial solitons, dissipative solitons, localized patterns, cavity solitons, or auto-solitons depending 
on the physical contexts in which their were observed. 
Localized structures can occur either in the presence \cite{Pomeau} or in the absence \cite{Nonturing} of a  symmetry breaking instability. 
In the last case, bistability between uniform states is a prerequisite condition for LS's formation. 
However, in the presence of symmetry breaking instability, the coexistence between a single uniform solution and a patterned state allows for the stabilization of LS's \cite{Nonturing}. 
In this case, bistability condition  between uniform solutions is not a necessary condition for generating LS's  \cite{Residori2009}.

\begin{figure}[t]
\centering
\includegraphics[width=8.0 cm]{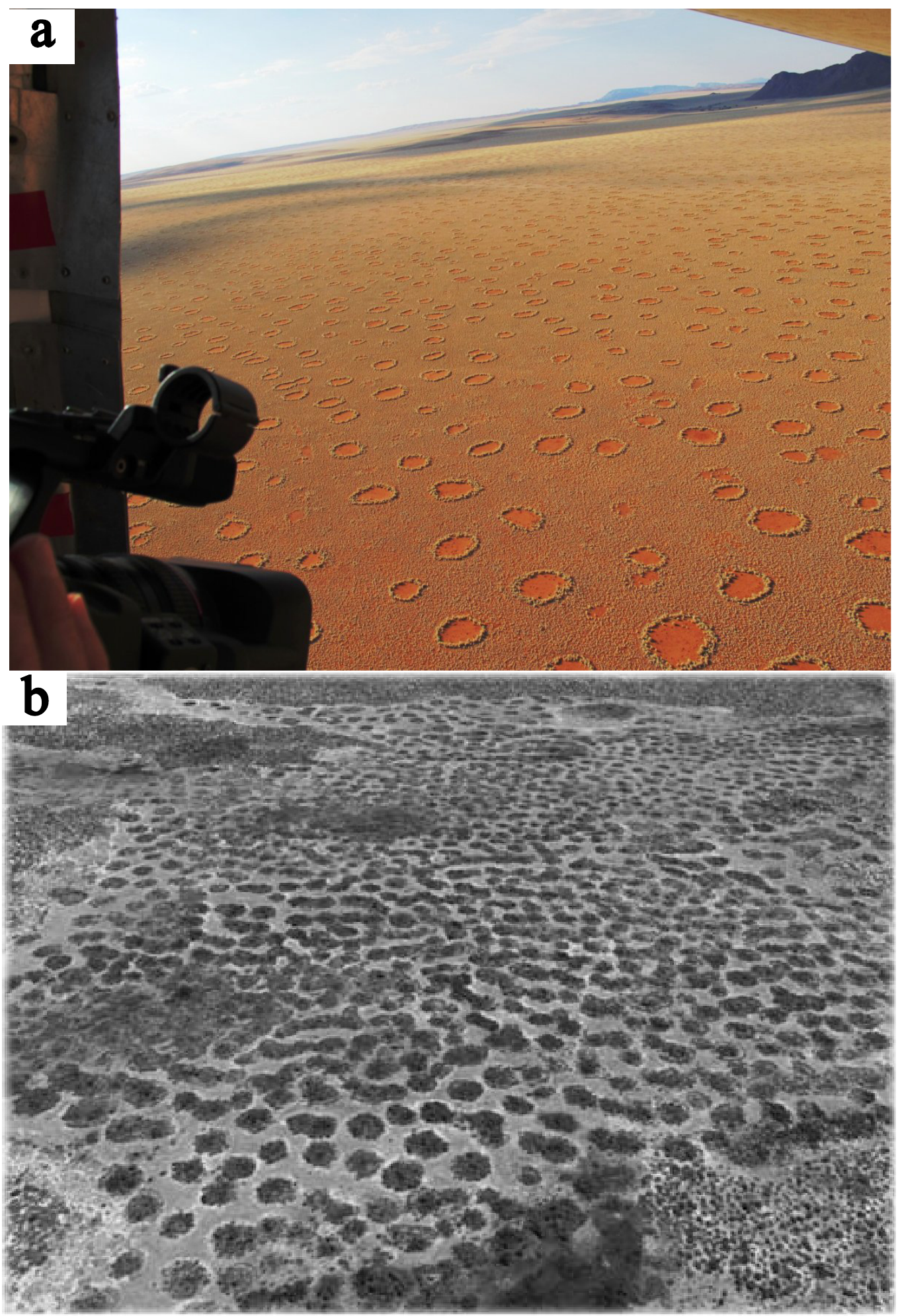}
\caption{(Color online) Localized vegetation patterns: (a) 
An aerial photo taken on 05 March 2010 shows fairy circles in the 
Namibrand region, Namibia. Image courtesy of Professor Norbert Juergens. 
(b) An aerial photograph showing localized  vegetation patches,  Zambia \cite{Borgogno}.
}
\label{Picture}
\end{figure}

Spatial coupling in many spatially extended systems is nonlocal. The Kernel function that characterizes the nonlocality  
can be either weak and strong. If the Kernel function decays asymptotically to infinity slower than an exponential 
function, the nonlocal coupling is said to be strong \cite{Escaff11}.  
If the Kernel function decays asymptotically to infinity faster than an exponential function, the nonlocal coupling is said to be weak. 
Self-organization phenomenon leading to the formation of  either extended or localized patterns under a local and nonlocal  coupling occur in various systems such as  fluid dynamics \cite{Kolodner}, firing of cells \cite{Hernandez-Garcia,Murry}, propagation of infectious diseases \cite{Ruan}, chemical reactions \cite{Kuramoto,Shima}, population dynamics \cite{Clerc2005,CEK3,CEK4}, nonlinear optics \cite{Krolikowski1,Krolikowski2,Krolikowski3,Mihalache,Mihalache1,Mihalache2,Gelens07}, granular \cite{Aranson}, and vegetation patterns \cite{Lefever97,LT,Tlidi08,GIrald,Couteron}.

We focus on bistable regime far from any symmetry breaking instability, i.e. far from any Turing instability. In this case, the behavior of many systems is governed by front dynamics or domains between the homogeneous steady states.  When the nonlocal coupling is weak, the interaction between two fronts is usually described by the behavior of the tail of fronts. However, for strong nonlocal coupling, the interaction is controlled by the whole Kernel function and not by the asymptotic behavior of the front tails \cite{Escaff11}.

When  a non-local coupling is weak, the asymptotic behavior of front solutions is characterized by exponential decay or damping oscillations. In the former case, front interaction is always attractive and decays exponentially with the distance between the fronts.  In two dimensional setting, LS's resulting from fronts interaction are unstable. In the case of damping oscillations, front interactions alternate between attractive and repulsive with an intensity that decays exponentially with the distance between the two fronts \cite{saarloos}. For a fixed value of parameters, a family of stable one dimensional localized structures with different sizes has been reported in \cite{Clerc2005,Gelens1,Gelens2,Gelens3}. An important difference appears when considering a strong nonlocal coupling. 
This difference is that the interaction between fronts can be repulsive \cite{OtoPRL,OtoPTRS}. 

Strong nonlocal coupling has been observed experimentally in various systems. Indeed, several experimental measurements of  nonlocal response in the form of Lorentzian or a generalized Lorentzian have been carried out in nematic liquid crystals cells  \cite{Hutsebaut,Henninot}. Experimental reconstruction of strong nonlocal coupling has also been performed in photorefractive materials \cite{ Minovich}. In this case, the strong nonlocal coupling is originating from the thermal medium effects. In population dynamics such as vegetation,  it has been shown experimentally that seed dispersion may be described by a  Lorentzian \cite{Howe}. 

We consider a proto model for population dynamics, namely the strong nonlocal  Nagumo equation \cite{OtoPRL}. 
This model possesses two relevant properties: strong nonlocal coupling and bistability between uniformly vegetated and bare states.  
We focus on a regime far from any symmetry breaking  or Turing type of instabilities. 
In this regime localized structures resulting from strong nonlocal coupling can be stabilized in a wide range of parameters \cite{OtoPRL}.  
We will investigate two types of localized vegetation structures: i) isolated or randomly distributed circular areas devoid of any vegetation, often called fairy circles (FC's), and ii) isolated or randomly distributed circular areas of vegetation, surrounded by a bare region.
An example of FC's is shown in an areal photograph (see Fig. 1a). 
They are observed in vast territories in southern Angola, Namibia, and South Africa \cite{Fraley,Albrecht,Getzin,Cramer2013,Picker12,Jankowitz,Naude,Jorgen,Grube}. 
The size of these circles can reach diameters of up to 12m.
An in-depth investigation of several hypotheses concerning their origin has been performed by van Rooyen et al. \cite{Rooyen}. 
In this study, these authors have been able to excluded the possible existence  radioactive areas inapt for the development of plants, the termites activity and the release of allelopathic compounds. 
We attribute two main ingredients to their stabilization: the bistability between the bare state and the uniformly vegetation state, and Lorentzian-like  non-local coupling that models the competition between plants \cite{OtoPTRS}. 
We provide detailed analysis of the  fairy circle formation in a simple population dynamics model. In addition we show that the above mechanism applies to another type of localized structures that consist of isolated or randomly distributed vegetation patches surrounded by a bare state. 
An example of this behavior is shown in Fig. 1b. 
In this paper, we investigate analytically and numerically the formation of both---fairy circles and localized patches--and their existence range. 
Our theoretical analysis  shows that exist a Maxwell point above which localized patches are stable, while below this point fairy circles appear. 
Finally, we investigate how the degree of aridity affects the width of both types of  localized vegetation structures.

This paper is organized as follows. After a briefly introducing the model  describing the vegetation dynamics, namely the Nagumo model with strong nonlocal coupling mediated by a  Lorentzian function (Sec. II). We  describe the dynamics of a single front in one dimension, and its asymptotic behaviors (Sec. III). The analytical and numerical analysis of the interaction between fronts connecting the uniformly vegetated and the bare steady states is described in Sec. IV, where we discuss the formation of both fairy circles and localized patches.   Close to the Maxwell point, we derive a formula for the width of  both localized structures as a function of the degree of the aridity in one dimension.  We conclude in Sec. V.

\section{The Nagumo model}
Several models describing vegetation patterns and self-organization in arid and semiarid landscapes have been proposed
during last two decades. They can be classified into three types.  
The  first approach is based  on the relationship between the structure of individual plants and the facilitation-competition 
interactions existing within plant communities \cite{Lefever97,LT,Tlidi08,Lefever-Turner,Martinez-Garcia}. 
The second is based on the reaction-diffusion approach which takes into account of the influence  of water transport by below ground diffusion and/or above ground run-off \cite{Klausmeier,Meron0,HilleRisLambers,Okayasu,Sherat,Wang,Kefi}.  
The third  approach focuses on the role of environmental randomness as a source of noise induced symmetry breaking transitions \cite{Odorico1,Odorico2,Borgogno,Ridolfi}. 
Recently, the reduction of a generic interaction-redistribution  model,  which belong to the first class of ecological type of models  \cite{Tlidi08}, to a Nagumo-type model has been established \cite{OtoPTRS}. 
Here, we consider the variational nonlocal Nagumo-type equation
\begin{equation}
\partial_{t}u=u(\alpha-u)(u-1)+\nabla^{2}u+\epsilon u\int_{\Omega} u^{2}({\bf{r}}+{\bf{r}}^{\prime},t)K({\bf{r}}')d{\bf{r}}'\label{NAGUMO}
\end{equation}
where $u\left({\bf{r}},t\right)$ is a normalized scalar field that represents the population density or biomass, $\alpha$ is a parameter describing the environment adversity or the degree of aridity, $t$ is time.
We consider that population or plant community established on a spatially uniform territory $\Omega$. 
The vegetation spatial propagation, via seed dispersion and/or other natural mechanisms, is usually modeled by a non-local coupling with a Gaussian like Kernel  \cite{Lefever97,LT,Tlidi08}.  
For simplicity, we consider only the first term in Taylor expansion of the dispersion. This approximation leads to the  Laplace operator, $\nabla^{2}=\partial_{xx}+\partial_{yy}$ acting in the space ${\bf{r}}=(x,y)$.  
The last term in Eq. \ref{NAGUMO} describes the competitive interaction between individual plants through their roots. The nonlocal coupling intensity, denoted by $\epsilon$, should be positive to ensure a competitive interaction between plants.  
The kernel function has the form $K\left(\bf{r}\right)=\delta\left(\bf{r}\right)-f_{\sigma}\left(\bf{r}\right)$, where $\delta\left(\bf{r}\right)$ is the delta function.   The inclusion of the $\delta$ function in the analysis allows us to avoid the variation of the spatially uniform states as a function  the nonlocal intensity $\epsilon$, and 
\begin{eqnarray}
f_{\sigma}\left({\bf{r}}\right)=\frac{N_{n}}{1+\left(|{\bf{r}}|/\sigma\right)^{n}},
\label{Lorentzian}
\end{eqnarray}
which has an effective range $\sigma$. 
For the sake of  simplicity, we consider  that the length $\sigma$ is a 
constant, independent of the biomass.  
Hence, we assume all plants have the same root size, that is, we neglect  allometric effect \cite{Leferver09,Lefever-Turner}.
At large distance, the asymptotic behavior of the Kernel is determined by $n$, and $N_n$ is normalization constant. 

 It is worth to emphasize that, since the strong nonlocal term in Eq. (\ref{NAGUMO}) models the interaction between individual plants at the community level , we refer to it as strong nonlocal interaction or coupling. 
In contrast, non-localities describing transport processes, as seed dispersion  \cite{Lefever97,LT,Tlidi08}, for sake of simplicity, we are modeling by the Laplace operator.

The  Eq. (\ref{NAGUMO}) is variational and it is described by 
\begin{equation}
\partial_{t}u= -\frac{\delta\mathcal{F}}{\delta u} \Rightarrow \frac{d\mathcal{F}}{dt} \leq 0,  \nonumber
\end{equation}
where $\mathcal{F}$  is a Lyaponov functional that can only decrease in the course of time. Accordingly, 
any initial distribution  $u\left(\vec{r},t\right)$ evolves towards a homogeneous or inhomogeneous (periodic or localized) 
state corresponding to a local or global minimum of $\mathcal{F}$. The  Lyaponov functional  reads
\begin{eqnarray}
\mathcal{F}[u]={\int}_{\Omega}\left\{  \frac{1}{2}(\left|\nabla u\right|^{2}+V\left(u\right) \right\}  dr  \nonumber \\ 
+\frac{\epsilon}{4}{\int}_{\Omega}{\int}_{\Omega}u^{2}({\bf{r}})u^{2}({\bf{r}}^{\prime})K\left({\bf{r}}-{\bf{r}}^{\prime}\right) drdr^{\prime} \nonumber  \label{EnergiaLibre}%
\end{eqnarray}
and
\begin{equation}
V\left(u\right) = \frac{u^2}{4}\left(u-1\right)^2 + \frac{u^2}{6}\left(\alpha - 1/2\right)\left(3-2u\right).
\label{Potencial}
\end{equation}

The  Eq. (\ref{NAGUMO}) admits  three spatially uniform solutions $u=0$, $u=\alpha$ and $u=1$. The bare state $u=0$ is 
always stable, and represents not plant state. The uniform state $u=\alpha$ is always unstable. For large values of $\alpha$, 
the climate becomes more and more arid. 
The  uniformly vegetated state $u=1$ may undergo a may undergo a symmetry breaking type of
instability (often called  Turing  instability),  that leads to pattern formation. 
In one dimensional system, and for  $n=2$, the threshold for that instability satisfies
 \begin{equation}
\beta = \epsilon\sigma^2\exp(-\beta), 
\label{EqInestabilidad}
\end{equation}
with $\beta=\sqrt{1+\sigma^2(2\epsilon + \alpha -1)}-1$. 
In what follows, we will focus on a regime far from any pattern forming instability. 

\section{Fronts}
{

We consider a bistable regime where $u=0$ and  $u=1$ are both linearly stable. 
From Eq. (\ref{Potencial}), $V\left(0\right)=0$ and
$V\left(1\right)=\left(\alpha - 1/2\right)/6$. 
Therefore, when $\alpha<1/2$, the most favorable state is the uniformly vegetated one. When $\alpha>1/2$, the bare state is more stable than the uniformly vegetated one.  There exist a particular point where both states are equally stable. This point is usually called the Maxwell point \cite{Goldstein}, and corresponds to  $\alpha=1/2$.

Depending on the value of  $\alpha$,  front connecting both state will propagate towards the most stable state.  An example of a single front is illustrated  in Fig. \ref{fig_VelFront}.a. The time-space diagram of Fig. \ref{fig_VelFront}.b shows  how the most stable state, corresponding to the bare state ($\alpha>1/2$), invades  the uniformly vegetated state,  with a constant speed.
Fronts propagate following the minimization of the potential (\ref{EnergiaLibre}) and the  front velocity is proportional to the energy difference between equilibria $V\left(1\right)-V\left(0\right)$.  The front is motionless at the Maxwell point for  $\alpha=1/2$. At this point and in the absence of nonlocal coupling, $\epsilon=0$, front solutions read
\begin{equation}
u_{\pm}\left(x-x_0 \right) = \frac{1}{2}\left(1 \pm \tanh\left(\frac{x-x_0}{2\sqrt{2}}\right) \right),\label{KINK}
\end{equation}
where $x_0$ corresponds to the interphase position. 
The front $u_+$ links the barren state from $x=-\infty$ to the uniformly vegetated state at $x=\infty$. 
The opposite connection correspond to $u_-$.

The asymptotic behavior of the  front solutions (\ref{KINK}) obeys to  an exponential law in the form
\begin{eqnarray}
 u_+\left(x\ll x_0\right)\approx e^{\left(x - x_0\right)/\sqrt{2}}
\nonumber, \\ 
 u_+\left(x\gg x_0\right)\approx 1- e^{-\left(x - x_0\right)/\sqrt{2}}
.\nonumber
\end{eqnarray}

\begin{figure}[tbp]	
\includegraphics[width=8cm]{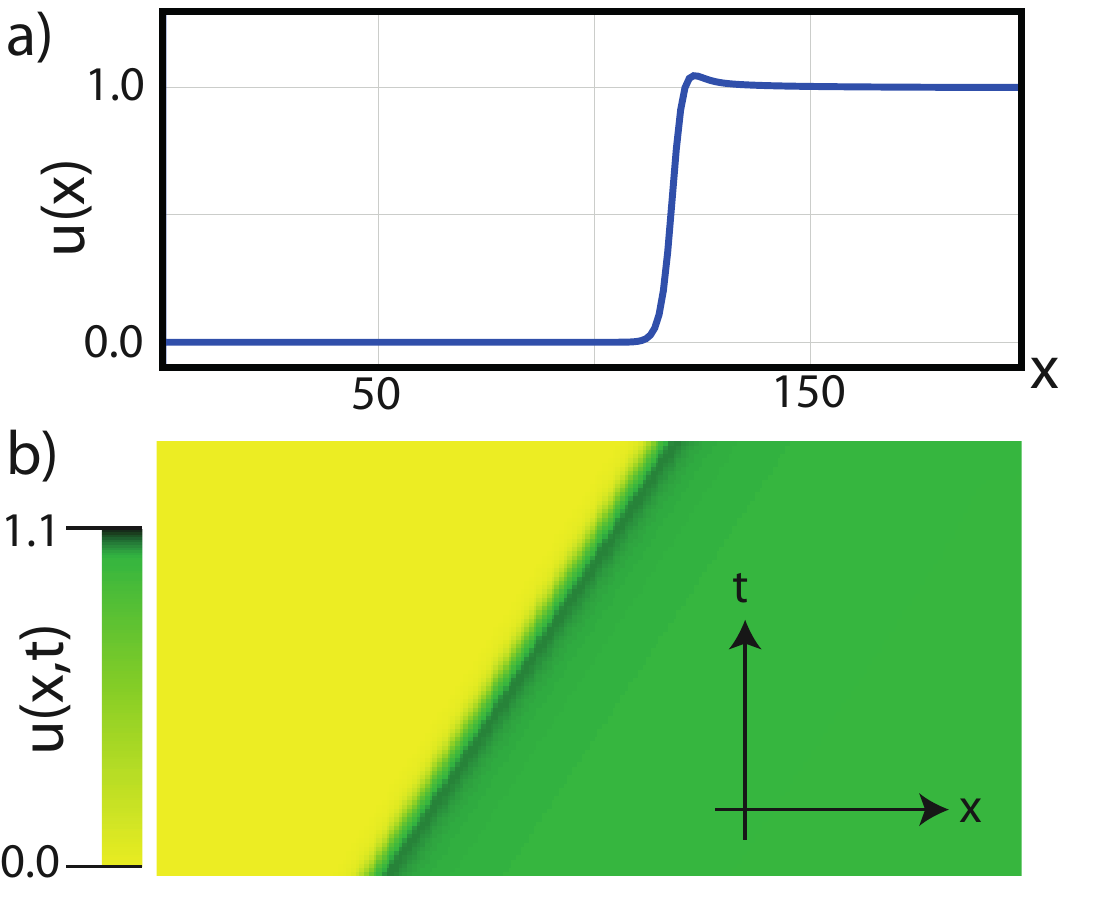}
\caption{(Color online) Front propagation obtained from numerical simulations of Eq. (\ref{NAGUMO}). 
(a) Biomass front profile. 
(b) Spatiotemporal diagram that shows the front movement with a constant speed. 
The parameters are $\alpha=0.51$, $\sigma=0.7$, $n=2$ and $\epsilon=1.0$.}
\label{fig_VelFront}
\end{figure}

Around the barren state ($u=0$), the inclusion of the nonlocal interaction does not modify the asymptotic behavior of 
the front, since the nonlocality in model (\ref{NAGUMO}) is  nonlinear. Let us examine the effect of nonlocal term around 
the  uniformly vegetated steady state ($u=1)$. For this purpose,  let us assume   that the asymptotic behavior of the front obeys to an exponential law in the form
\begin{eqnarray}
 u\left(x\gg x_0\right)\approx 1- ce^{-\lambda\left(x - x_0\right)},
\nonumber
\end{eqnarray}
where $c$ is a constant, and the exponent $\lambda$ obeys the equation
\begin{eqnarray}
 \lambda^2 - \frac{1}{2} + \epsilon\left(3 - 2 g_{\sigma}\left(\lambda\right) \right) = 0,
\nonumber
\end{eqnarray}
where
\begin{eqnarray}
g_{\sigma}\left(\lambda\right)  = \int_{-\infty}^{\infty} \cosh\left(\lambda x\right)f_{\sigma}\left(x\right) dx.
\nonumber
\end{eqnarray}
This equation has been successfully used to explain the emergence of localized domains. 
When $\lambda$-solutions have non-null imaginary part, spatially damped oscillations on the front profile are induced,  leading to the stabilization of localized domain.  This mechanism is well documented for either local or nonlocal  systems \cite{Clerc2005,Gelens1,Gelens2,Coullet2002}.

The $g_{\sigma}$ function exists  when the Kernel decays faster than an exponential one, i.e., a weak  nonlocal coupling.
In the case of strong nonlocal interaction, $g_{\sigma}$ diverges and the above analysis is no more  longer valid. 

To determine the asymptotic behavior of the front around the uniformly vegetated state under strong nonlocal coupling, we perform a regular perturbation analysis in terms of small parameter $\epsilon$. At the Maxwell point, we expand the field as
\begin{eqnarray}
\label{Expansion}
u\left(x\right) = u_0\left(x\right) + \epsilon u_1\left(x\right) + \epsilon^2 u_2\left(x\right) + ...,
\end{eqnarray}
where $u_0=u_+$ is the motionless front provided by the Eq. (\ref{KINK}). 
Replacing (\ref{Expansion}) in Eq.(\ref{NAGUMO}), and making an expansion in series of $\epsilon $.
At order $\epsilon $, we obtain:
\begin{eqnarray}
&\left\{\partial_{xx} - 1/2 + 3u_0\left(1-u_0\right)\right\} u_1 = 
\nonumber \\ 
&\left(u_0\int_{-\infty}^{\infty} u_{0}^{2}(x+x^{\prime})f_{\sigma}(x')dx'-u_0^3\right).\nonumber
\end{eqnarray}
Let us focus in the region $x\gg x_0$, and neglecting all exponential corrections coming from $u_0$,  then we obtain
\begin{equation}
\left\{\partial_{xx} - 1/2 \right\} u_1 = -\int_{x-x_0}^{\infty} f_{\sigma}\left(x'\right) dx'.\label{asymptotic}
\end{equation} 

\begin{figure}[tbp]
\includegraphics[width=8cm]{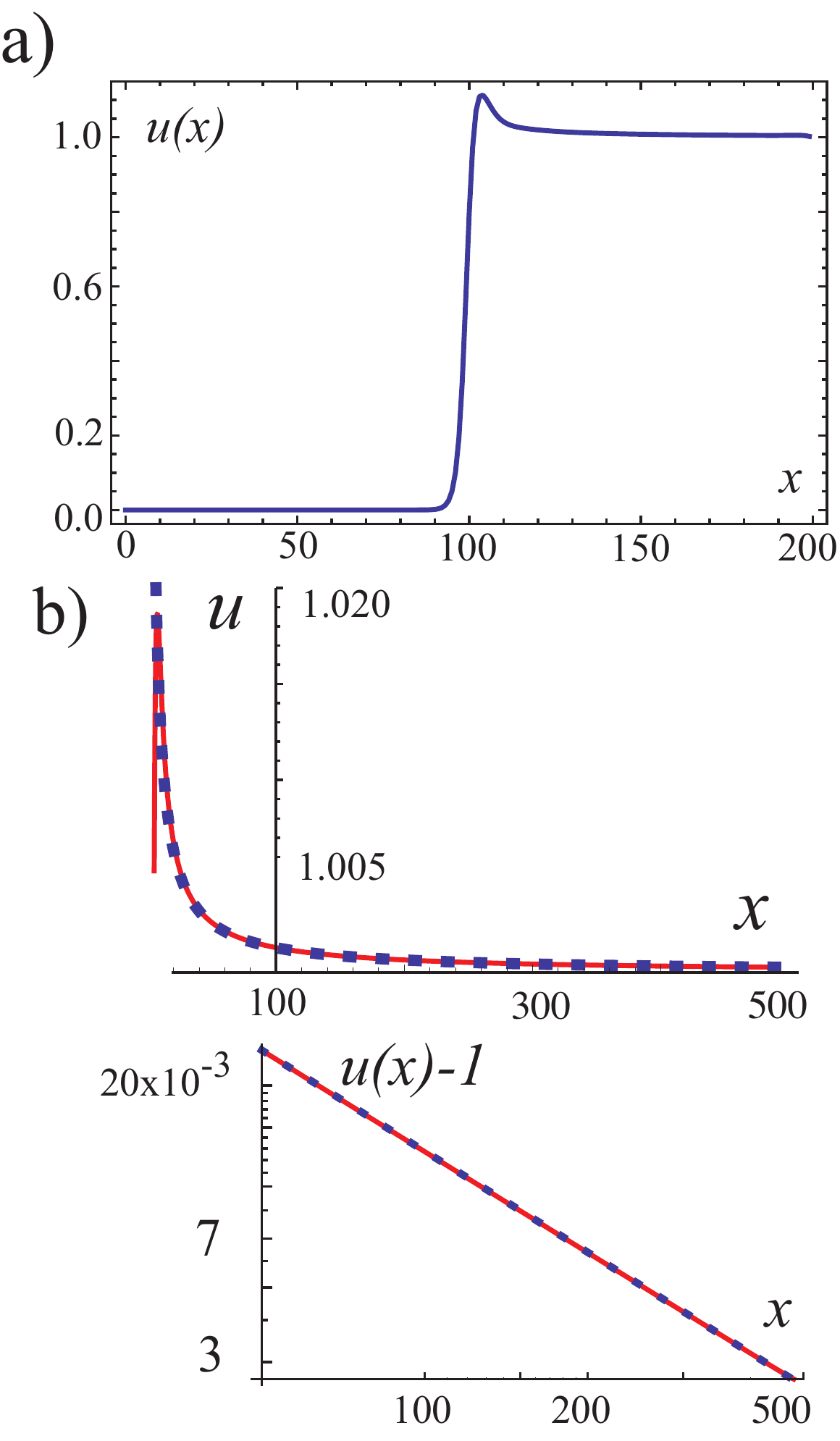}
\caption{(Color online) Numerical solutions of model (\ref{NAGUMO}), with the Lorentzian kernel (\ref{Lorentzian}). 
(a) Front profile at the Maxwell point $\alpha=0.5$, other parameters are $\sigma=2$, $n=2$ and $\epsilon=0.35$. 
(b) Decay of front to the uniformly vegetated state for $\alpha=0.5$, 
$\sigma=2$, $n=2$ and $\epsilon=0.1$. In (b), the upper panel shows the comparison between the numerical calculation (dotted line) and the analytical estimation (\ref{Frontstrong})
(continuous line), while the lower panel displays the same comparison in
a log-log plot.}
\label{fig_ProfileFront}
\end{figure}

Equation (\ref{asymptotic}) is a linear inhomogeneous equation for the correction $u_1$. 
For strong nonlocal coupling, the particular solution of Eq. \ref{asymptotic} dominates over the homogeneous one which is exponentially small.  For instance, if we consider a Kernel like
\begin{eqnarray}
f_{\sigma}\left(x\right)\approx \frac{N}{x^{n}} \text{ }\text{ for }\text{ }x\gg 1,
\label{power}
\end{eqnarray}
with $n >1$ and $N$ a normalization constant. For $x-x_0\gg 1$, the front approaches asymptotically the following solution
\begin{equation}
u \approx  1 + \frac{2\epsilon N}{\left(n -1\right)\left(x-x_0\right)^{n-1}} .
\label{Frontstrong}
\end{equation}
This solution decays according to a power law $1-n$. To check the power obtained from the above analysis, we perform a numerical simulations of the full model Eq. (\ref{NAGUMO}).  The  result obtained from Eq. (\ref{power}) and the numerical simulations are shown in in  Fig. \ref{fig_ProfileFront}. Both results agree perfectly  without any adjusting parameter.  Note that both analytical calculations and numerical simulations predict the existence of one peak in the spatial profile of the front. This peak takes place at the interface separating both homogeneous steady states as shown in Fig.~\ref{fig_ProfileFront}.

\section{Localized vegetation patterns}

Far from a symmetry breaking instability, localized structures can be stable as a results of front interactions. 
This phenomenon occurs  when the spatial profile of the front exhibits damped oscillations 
\cite{Clerc2005,Coullet2002}. 
However,  around the bare state damped oscillations are nonphysical since the biomass is a positive defined quantity. A stabilization mechanism of LS's  based on combined influence of strong nonlocal coupling and bistability has been proposed \cite{OtoPRL}. This mechanism has been applied to explain the origin of the fairy circles phenomenon in realistic ecological model \cite{OtoPTRS}. To the best our knowledge, there is no other analytical understanding of the circular shape of the fairy circle. We have shown in addition that the diameter of the single fairy circle is intrinsic to the dynamics of the system such as the competitive interaction between plant and the redistribution of resources \cite{OtoPTRS}. We believe that extrinsic causes such as termite or ant or others external environment cannot explain the circular shape of  fairy circles.   

In this section  we provide a detailed analysis of front interactions leading to stabilization of both fairy circles and localized patches.  For both types of localized vegetation structures, we discuss how the level of aridity affects their diameter.

\begin{figure}[tbp]
\includegraphics[width=8cm]{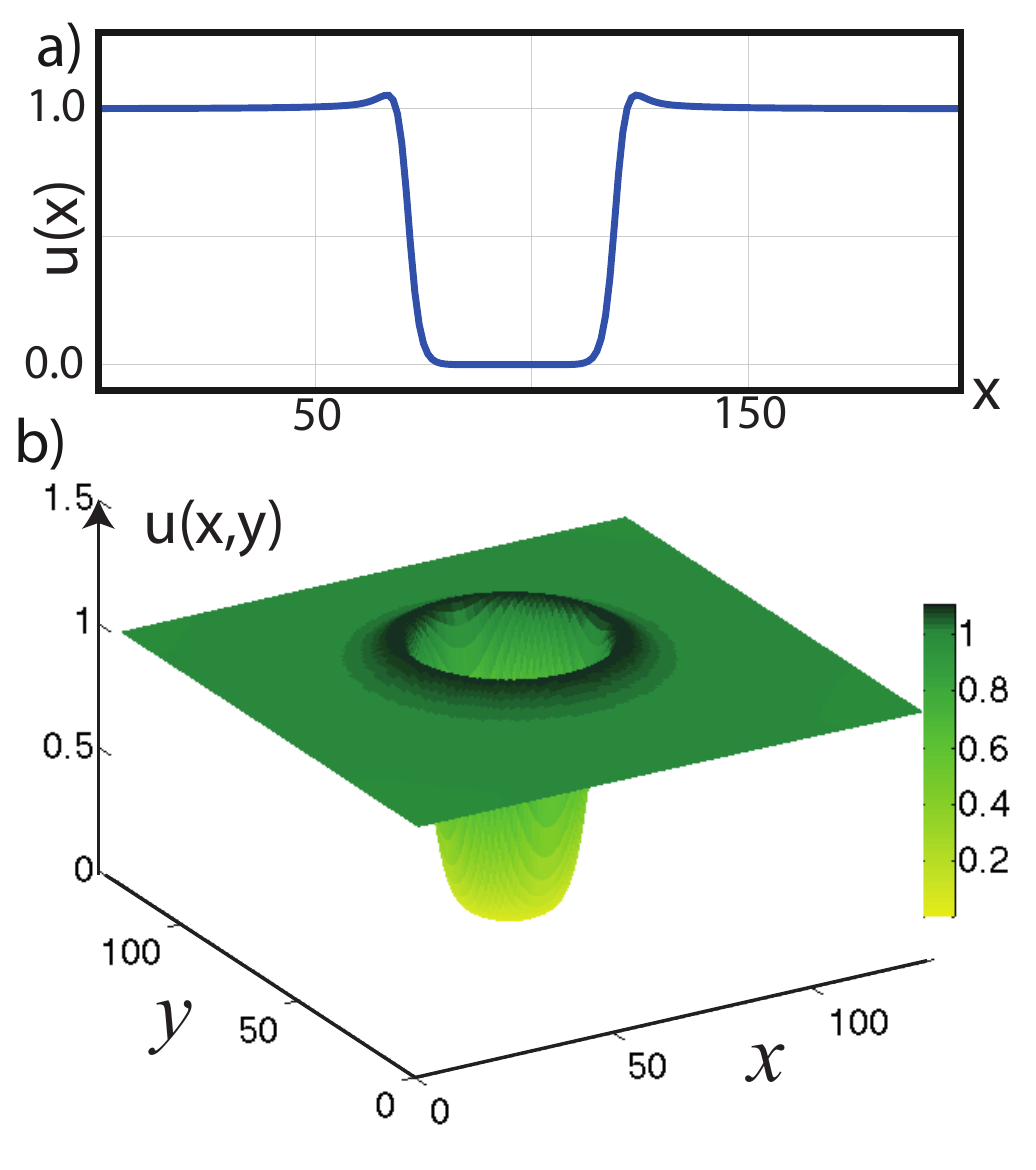}
\caption{(Color online)
Stationary fairy circle obtained from numerical simulations of Eq. (\ref{NAGUMO}). 
(a) One dimension with $\alpha=0.492$, $\sigma=0.8$, $n=2$ and $\epsilon=0.5$.  
(b) Two dimensions with $\alpha=0.38$, $\sigma=0.2$, $n=2.2$ and $\epsilon=1$.
}
\label{fig4}
\end{figure}

\subsection{Fairy circles}
{

The model Eq. (\ref{NAGUMO}) admits stable localized structures in the form of bare state embedded in herbaceous vegetation matrix. An example of such behavior is shows in  Fig. \ref{fig4}. They are stable and permanent structures. A single fairy circle exhibits a fringe formed by tall grass that separate the bare state to the uniformly vegetated  
as shown in Fig. \ref{Picture}a.  From numerical simulations, we see that the biomass possess one peak that takes place in between the bare and the uniformly vegetated states as shows in Fig. \ref{fig4}.  This can be explain by the fact that inside the circle the competition between plant is small. Indeed the length of the plant root is much smaller than the diameter of the fairy circle.

We analytically investigate the formation of  a single fairy circle in one spatial dimension. We focus on the parameter region near to Maxwell point ($\frac{1}{2}-\alpha=\eta\ll1$) and we consider a small non-local coupling ($\epsilon\ll1$).  We look for a solution of  Eq. (\ref{NAGUMO}) that  has the form of a slightly perturbed linear superposition of two fronts
\begin{eqnarray}
u(x,t)=u_+ \left(x-\delta(t)\right)+u_-\left(x+\delta(t)\right)+W,
\label{ansatz_bfc}
\end{eqnarray}
 where  $W(x,u_+,u_-)$, $\partial_{t}\delta$, $\eta$ and $\epsilon$ are small in a sense that will be pinned down bellow. While $u_{\pm}$ are defined by Eq. (\ref{KINK}).

Replacing anstatz (\ref{ansatz_bfc}) in Eq. (\ref{NAGUMO}) and neglecting high order terms in $\epsilon$, we obtain
\begin{eqnarray}
-(\partial_{x}u_{+}-\partial_{x}u_{-})\partial_{t}\delta  =u_{+}u_{-}(3-3u_{+}-3u_{-})  \nonumber \\
 +\epsilon u\int_{-\infty}^{\infty}u^{2}(x+x',t)K(x')dx'  \nonumber \\
 -\eta(u_+ + u_-- 1)(u_+ + u_-)  +LW+h.o.t.,
\label{Eq_Completa_Ansatz}
\end{eqnarray}
where the linear operator $L$ has the form
\[
L\equiv -\frac{1}{2}+3u_{+}-3u_{+}^{2}+\partial_{xx} +3u_{-}-3u_{-}^{2}-6u_{+}u_{-}.
\]
To solve the above equation, we consider the inner product $ \left\langle g|h\right\rangle \equiv \int_{-\infty}^{+\infty}(g(x)h(x))dx$. Then, 
the linear operator $L_+$ is self-adjoint and $L\partial_{x}u_{\pm}\approx 0$. The solvability condition gives
\begin{eqnarray}
-\left\langle \partial_{x}u_{+}|\partial_{x}u_{+}\right\rangle\partial_{t}\delta & = & G+\eta\left\langle \partial_{x}u_{+}|(1-u_+)u_+\right\rangle,
\label{Eq_Delta_G}
\end{eqnarray}
where
\begin{eqnarray*}
G(u_\pm,\sigma) & = & \left\langle \partial_{x}u_{+}|u_+\int_{-\infty}^{\infty}u^{2}(x',t)K(x'-x)dx'\right\rangle.
\end{eqnarray*}

We neglect also  the terms smaller than $1/{\delta^{2n-1}}$ in Eq. (\ref{Eq_Delta_G}). 
Then, it is obtained 
\begin{eqnarray}
\partial_{t}\delta=\frac{3\sqrt{2}\epsilon N_{n}\sigma^{n}}{(n-1)(2\delta)^{n-1}}-\sqrt{2}\eta,
\label{Eq_deltan}
\end{eqnarray}
Strictly speaking, this equation of motion is quantitatively valid when $\partial_{t}\delta\sim \epsilon/\delta^{n-1} \sim \eta\ll 1$.

\begin{figure}[tbp]
\includegraphics[width=8cm]{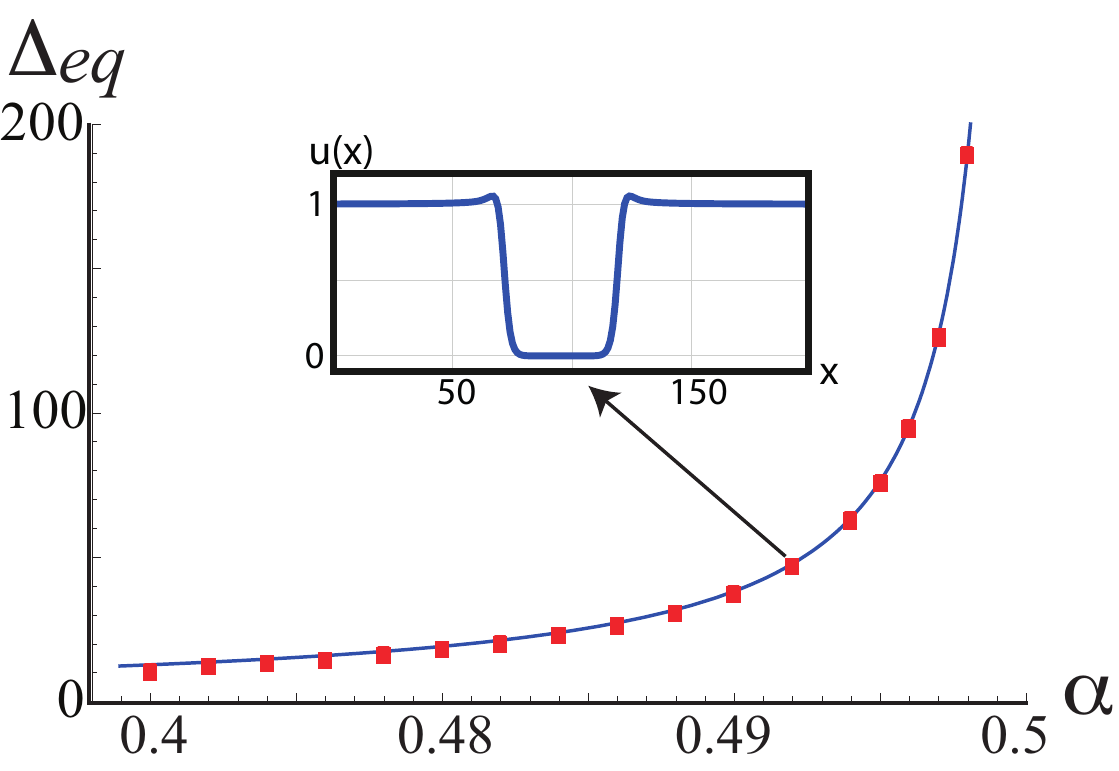}
\caption{(Color online) Width of a fairy circle, as a function of aridity $\alpha$ in one dimension. 
The  solid line represents the analytical prediction from Eq. (\ref{Eq_delta_size}). 
The squares are obtained by numerical simulations of Eq. (\ref{NAGUMO}). 
The inset is the spatial profile of biomass. 
Parameters are $\sigma=0.8$, $n=2$ and $\epsilon=0.5$. 
}
\label{fig6}
\end{figure}

This result is valid for any power $n>1$ in the Kernel function (\ref{Lorentzian}). We consider $n=2$ and $N_n=1/\pi\sigma$ 
in one dimension. Then, the equation (\ref{Eq_deltan}) reads
\begin{eqnarray}
\partial_{t}\Delta=\frac{6\sqrt{2}\epsilon\sigma}{\pi\Delta}-2\sqrt{2}\left(\frac{1}{2}-\alpha\right),
\label{Eq_Delta}
\end{eqnarray}
where $\Delta=2\delta$ is  the width of the LS.
The equilibrium width is given by
\begin{eqnarray}
\Delta_{eq}=\frac{3\epsilon\sigma}{\pi\left(\frac{1}{2}-\alpha\right)}.
\label{Eq_delta_size}
\end{eqnarray}
The linear stability analysis allows us to determine the eigenvalue  $\lambda=-2\sqrt{2}\pi\eta^2/3\epsilon\sigma$. Therefore, for competitive interaction, i.e., $\epsilon>0$, fairy circles are always stable. 

We plot the width of the FC in one dimension as function of the aridity $\alpha$ in  Figure \ref{fig6}. 
The width of localized structures grows as  aridity increases. 
At the Maxwell point, i.e., $\alpha=1/2$,  the width of the fairy circle  becomes infinite. 
In order to check the approximations used to derive the  equilibrium size (\ref{Eq_delta_size}), we perform numerical simulations of  Eq. (\ref{NAGUMO}). Both results are in good agreement, without any fitting parameter.

The fairy circles  are observed in vast territories in southern Angola, Namibia, and South Africa \cite{Getzin,Albrecht}, 
where the annual rainfall ranges between $50$ and $150$~mm  \cite{Rooyen}. The size of FC increases from South to 
North where the climate  becomes more and more arid \cite{Rooyen}. The size can also be affected by the rainfall and nutrients \cite{Cramer2013}.  
Fairy circles average diameters varies in the range of $2$ m-$12$ m \cite{Rooyen}.  In agreement with field observations,  
Fig.~\ref{fig6} shows indeed that the  fairy circles diameter increases with the aridity. 

Therefore, one-dimensional front interaction explains why the fairy circles size increase with the aridity. To wit, as environment aridity increases, the bare state becomes more and more favorable, increasing the fairy circles size. This mechanism demands, however, that the uniformly vegetated state must be always the most favorable one ($\alpha<1/2$), otherwise the bare state propagates indefinitely. The same tendency is observed in two-dimensional simulations \cite{OtoPTRS}.

}

\begin{figure}[b]
\includegraphics[width=8.5cm]{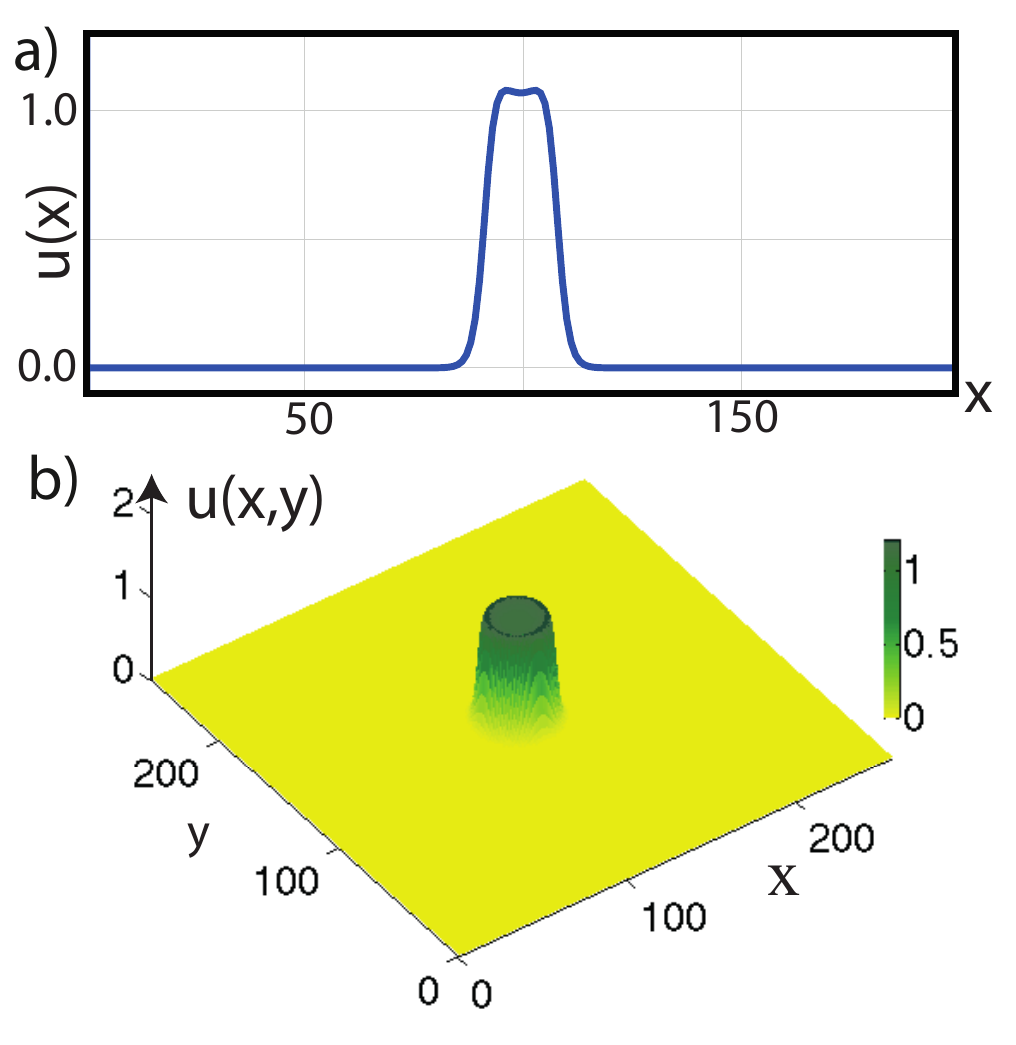}
\caption{(Color online)
Stationary localized patch obtained from numerical simulations of Eq. (\ref{NAGUMO}). 
(a) One dimension with $\alpha=0.52$, $\sigma=0.8$, $n=2$ and $\epsilon=0.5$.  
(b) Two dimensions with $\alpha=0.56$, $\sigma=0.1$, $n=2.5$ and $\epsilon=10$.
}
\label{fig7_IllusFFC}
\end{figure}

\subsection{Localized vegetation patch}

In this subsection, we investigate the formation of a single localized patch that consist of a circular vegetated state surrounded by a bare state. This behavior occurs for high values of the aridity parameter, i.e.,  $\alpha>1/2$.  An example of a single localized patch is illustrated in Fig. \ref{fig7_IllusFFC}. This localized solution corresponds to the counterpart of fairy circles.

Following a similar strategy than in the previous subsection (small $\epsilon$ and $\eta_f = \alpha - 1/2$),  the patch width $\Delta$ in one spatial dimension obeys \begin{eqnarray}
\partial_{t}\Delta=\frac{6\sqrt{2}\epsilon N_{n}\sigma^{n}}{(n-1)(\Delta)^{n-1}}-2\sqrt{2}\eta_f.
\label{Eq_deltanf}
\end{eqnarray}

\begin{figure}[tbp]
\includegraphics[width=8cm]{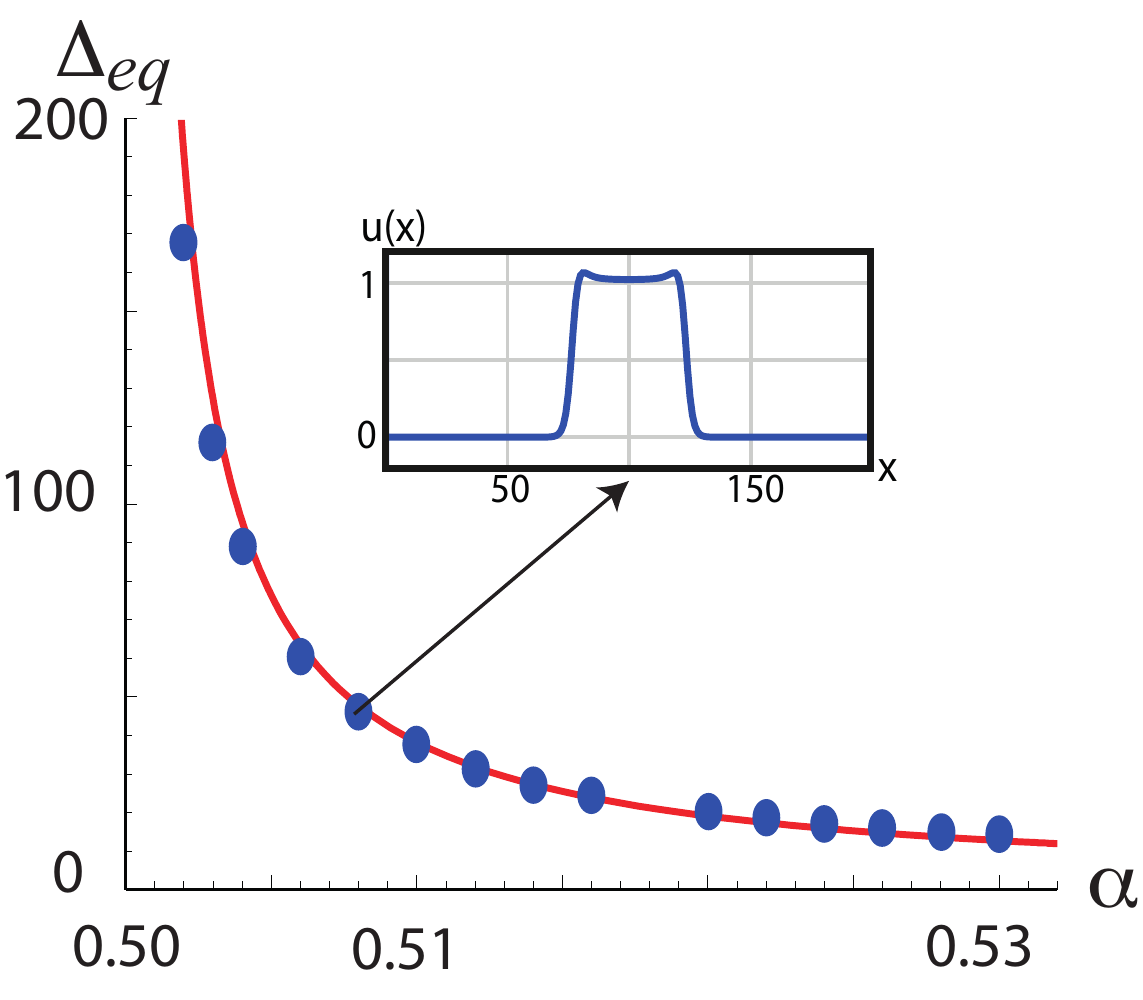}
\caption{(Color online) Width of a localized vegetation patch, as a function of aridity $\alpha$ in one dimension. 
The  solid line represents the analytical prediction from Eq. (\ref{Eq_Deltaf_size}). 
The  dots are obtained by numerical simulations of Eq. (\ref{NAGUMO}). 
The inset is the spatial profile of biomass. 
Parameters are $\sigma=0.8$, $n=2$ and $\epsilon=0.5$. 
}
\label{fig9}
\end{figure}
 
Note that, like in Eq. (\ref{Eq_deltan}), the result obtain in Eq. (\ref{Eq_deltanf}) is generic for any $n$ in (\ref{Lorentzian}). 
For $n=2$, the stable vegetated patch have the size
\begin{eqnarray}
\Delta_{eq}=\frac{3\epsilon\sigma}{\pi\left(\alpha-\frac{1}{2}\right)}.
\label{Eq_Deltaf_size}
\end{eqnarray}

The formula (\ref{Eq_Deltaf_size}) is plotted in Fig. \ref{fig9}  by a solid line. 
Confrontation with direct numerical computation for the localized patch width is in good agreement, as shown in Fig. \ref{fig9}. There is no available data from the field observation to confirm this theoretical prediction. 

\begin{figure}[tbp]
\includegraphics[width=8cm]{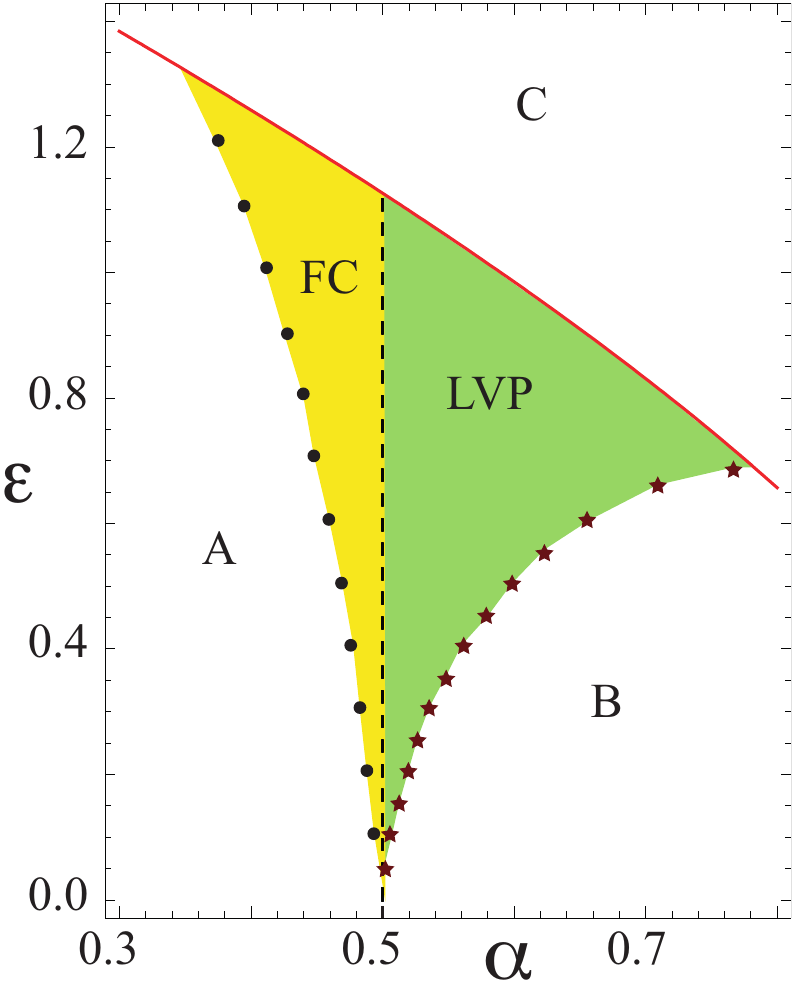}
\caption{(Color online) Bifurcation diagram of localized structures for model Eq. (\ref{NAGUMO}), 
in  parameter plane $(\alpha,\epsilon)$,
with $\sigma=0.8$ and $n=2$.
}
\label{fig5}
\end{figure}

\subsection{Bifurcation diagram}

In this subsection, we establish the bifurcation diagram for both types of localized vegetation structures. 
We fix the length of the completion between plant $\sigma$, and we vary the degree of aridity $\alpha$ and the strength of the competitive interaction $\epsilon$.  
We numerically establish a stability range of a single fairy circle and the localized patch in one dimensional setting. 
This analysis is summarized in the parameter plane $(\alpha,\epsilon)$ of Fig. \ref{fig5}. 
For $\alpha<1/2$, a single fairy circle is stable in the region FC as indicated in Fig.\ref{fig5}. 
This stability region is bounded from below by dots and bounded from left by the Maxwell point ($\alpha=1/2$). 
Dynamically speaking, dots correspond to a saddle-node bifurcation. 
The parameter zone A indicates the regime where a FC shrinks and disappears. 
For large values of the strength of the competition $\epsilon$, the uniformly vegetated state becomes unstable with respect to symmetry breaking instability. 
The threshold associated with this instability is represented by a  solid line. 
This line is obtained by plotting formula Eq. (\ref{EqInestabilidad}). 
This spatial instability impedes the existence of fairy circles in the region C, and may allow for the formation of periodic structures. 
For $\alpha>1/2$, a single fairy grows until infinity and disappears, and a localized vegetation patch appears. 
This structure is stable in the region  LVP, as shown in the bifurcation diagram of Fig. \ref{fig5}. 
The region B corresponds to a high degree of aridity. In this zone of parameters, a localized patch shrinks and disappears, and transition toward a bare state occurs.

\begin{figure}[t]
\includegraphics[width=8cm]{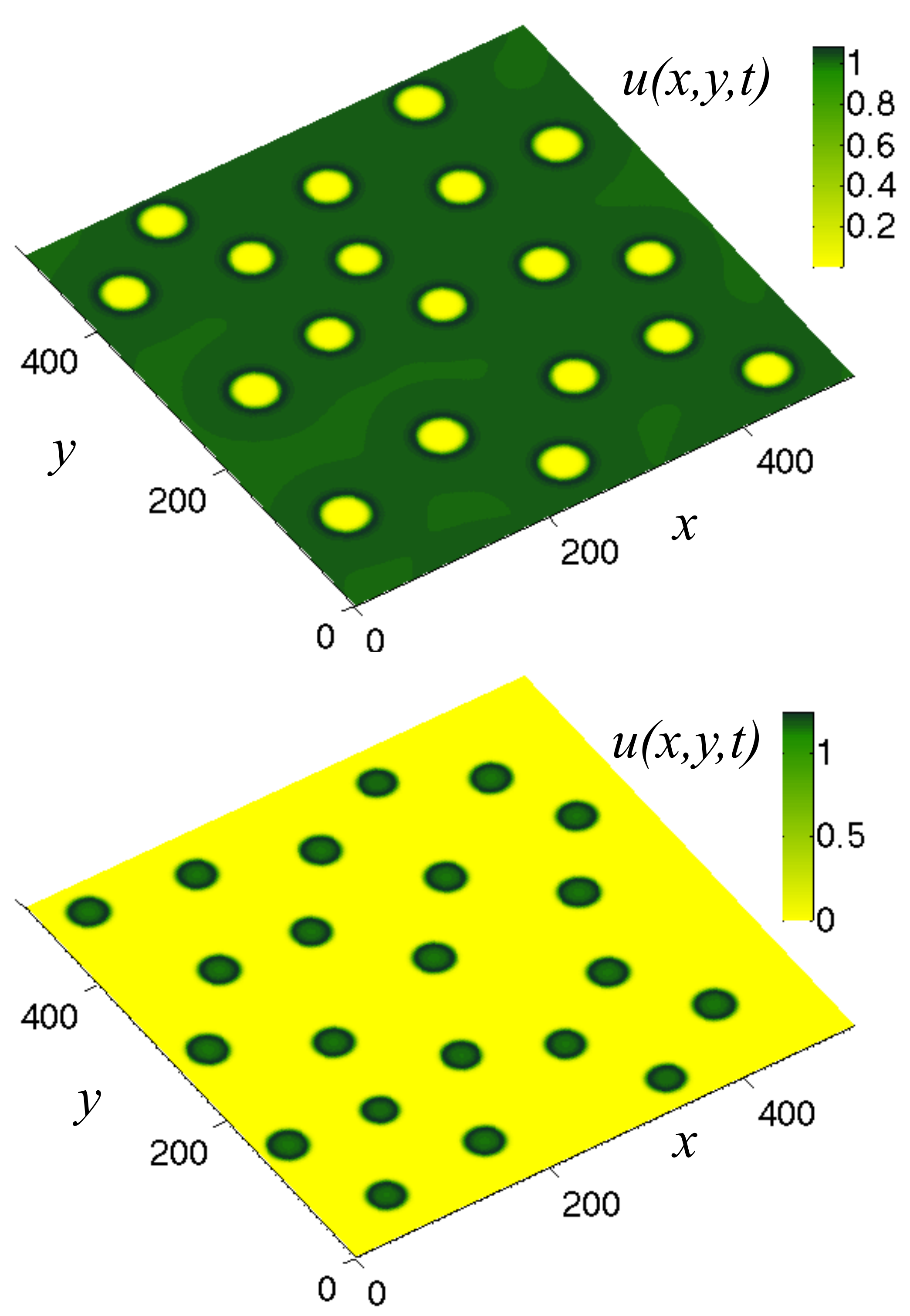}
\caption{(Color online) Two dimensional localized vegetation patterns 
obtained from numerical simulations of Eq. (\ref{NAGUMO}). (a) Multi-fairy circles for $\alpha=0.44$ (b) Multi-localized vegetation patches for  $\alpha=0.57$. Other parameters are $\sigma=0.1$, 
$\epsilon=10$ and $n=2.5$. 
}
\label{figEND}
\end{figure}

\section{Conclusions}

We have investigated the role of a strong nonlocal coupling in a bistable model namely the Nagumo model. 
This prototype model of population dynamics could be applied to vegetation dynamics.  
We have shown that far from any symmetry breaking or Turing type of instability, localized vegetation structures can be stabilized in large values of the aridity parameter. 
Their formation is attributed to the interaction between fronts mediated by a strong nonlocal coupling in the form of a Lorentzian.  
We have identified the following scenario: when increasing the level of the aridity, a large dip embedded in a uniformly vegetated state is formed. 
This structure has a single fringe peak that appeared in the spatial profile of the biomass. 
We have interpreted this behavior as a fairy circle. 
When increasing further the degree of aridity, localized vegetation patch can be formed in the system. 
This structure has a peak surrounded by the bare state. 
The localized structures reported in this work have a varying width as a function of aridity. 
In contrast, the width of localized vegetation structures found close to the symmetry breaking instability is determined by the most unstable Turing wavelength \cite{Tlidi08,Getzin02}.
We have established analytically a formula for the width of fairy circles and localized vegetation patches as a function the degree of aridity. 
The width of these localized structures is intrinsic to the dynamics of arid ecosystem and it is independent of external environmental effects, such as termites or ants. 
The results of  direct numerical simulations of model Eq. (\ref{NAGUMO}) agreed with the analytical findings.

In this paper we have focused our analysis on a single localized structure, several of them could be stable as shown in 
the Fig. \ref{figEND}. The formation of multi-dips or peaks localized structures, their interactions and their stability are 
under investigation. Understanding the formation of localized structures is central not only in arid ecosystems but also 
in spatially extended out of equilibrium systems.

\begin{acknowledgments}

M.G.C. acknowledges the financial support of FONDECYT project N$^o$ 1120320. D.E. acknowledges the 
financial support of FONDECYT project N$^o$ 1140128. C.F.-O. acknowledges the 
financial support of Becas Chile. M.T. received support from the Fonds National de la Recherche Scientifique (Belgium).

\end{acknowledgments}

\end{document}